\documentclass[twocolumn]{aastex61}

\usepackage{amsmath}          

\received{July 3, 2017}
\revised{September 13, 2017}
\accepted{October 6, 2017}

\submitjournal{ApJS}

\shorttitle{Terahertz spectroscopy of ND$_{2}$}
\shortauthors{Melosso et al.}

\begin{document}

\title{Terahertz spectroscopy and global analysis of the rotational spectrum of doubly deuterated Amidogen radical ND$_{2}$}

\author{Mattia {Melosso}}
\author{Claudio Degli Esposti}
\author{Luca Dore}
\affiliation{Dipartimento di Chimica "Giacomo Ciamician", Universit\`a di Bologna, via Selmi 2, 40126 Bologna, Italy}

\correspondingauthor{Mattia Melosso}
\email{mattia.melosso2@unibo.it}

\begin{abstract}
The deuteration mechanism of molecules in the interstellar medium (ISM) is still being debated. Observations of deuterium-bearing
 species in several astronomical sources represent a powerful tool to improve our understanding of the interstellar chemistry.
 The doubly deuterated form of the astrophysically interesting Amidogen radical could be a target of detection in space.
 In this work, the rotational spectrum of the ND$_{2}$ radical in its ground vibrational and electronic $X^{2}B_{1}$ states has been 
 investigated between 588 and 1131 GHz using a frequency modulation millimeter/submillimeter-wave spectrometer.
 The ND$_{2}$ has been produced in a free-space glass absorption cell by discharging a mixture of ND$_{3}$ and Ar.
 Sixty-four new transition frequencies involving $J$ values from 2 to 5 and $K_{a}$ values from 0 to 4 have been measured.
 A global analysis including all the previous field-free pure rotational data has been performed, allowing  for a more precise 
 determination of a very large number of spectroscopic parameters.
Accurate predictions of rotational transition frequencies of ND$_{2}$ are now available from a few GHz up to several THz.
\end{abstract}


\section{Introduction}
Recent discoveries of multiply deuterated molecules in the interstellar medium (ISM) \citep{Cec98, Lis02, Vas03, Par04} have confirmed 
the interest in deuterium-bearing species, since deuterium fractionation is a powerful tool to study the evolution of the Solar System
 \citep{Mil02, San02, Rou03}.
Both gas-phase models and active chemistry on grain-surface have been invoked to explain the formation of doubly or even triply 
deuterated species and their anomalous large amount \citep{Rob00, Cas02,Rob03}.
Further detections of multiply deuterated species are thus fundamental to constraint the chemical models and spectroscopic laboratory data are 
necessary to the goal.

The doubly deuterated Amidogen radical, ND$_{2}$, represents a possible target, since NH$_2$ is involved in the formation process 
of ammonia \citep{Her87} and all the deuterated isotopologues of ammonia have been detected in space 
\citep{Tur78, Rou00, Lis02}.
The main isotopologue NH$_{2}$ was first observed in 1943 in a comet \citep{Swi43} by optical astronomy.
More recently, it has also been detected via radio observation of the star-forming region Sagittarius B2 \citep{van93} and of many 
other sources \citep{Hil10, Per10}.
Moreover, in 2015 the rare $^{15}$N-substituted isotopologue has been observed through cometary emissions \citep{Rou14, Shi14}.
However, no detection of deuterated Amidogen, even in its mono deuterated form NHD, has been reported to date.

Nevertheless, the existence of deuterated forms of Amidogen in ISM seems to be plausible as predicted by the gas-phase chemical models, in which the dissociative recombination of partially deuterated intermediates results in a higher probability for the ejection of 
a  hydrogen atom than deuterium.
\citet{Rou05} presented a steady state model of the gas phase chemistry aimed at understanding the deuterium fractionation of ammonia in various sources: these authors considered different physical conditions and could derive a general trend.
At high density and high depletion, deuteration of N-containing species results very efficient at 10 K.
Actually, the abundances of singly and multiply deuterated
forms of H$_3^+$ can reach exceptionally high values under conditions of extreme CO-depletion, and
these high degrees of deuteration can propagate to the residual nitrogen-bearing species \citep{Flo06}, that appear to be less depleted
and subsist longer in the gas phase (see \citet{Ber97} and \citet{Bel04} for the cases of NH$_3$ and N$_2$H$^+$, respectively).
The deuterium fractionation is sensitive to the temperature, but it remains large for temperatures between 5 and 20 K.
Thus, the model predicts high fractional abundances of ammonia progenitors and their deuterated isotopologues, namely NH, ND, NH$_{2}$, 
NHD and ND$_{2}$, in dense cores.
In general, the predicted abundance ratios of the deuterated ammonia isotopologues and their progenitors agree reasonably well with existing observations, and, noticeably, the fractional abundance of ND$_{2}$ is estimated to be from 2 to 9 times larger than the ND$_3$ abundance.

The ND$_{2}$ radical has been long studied with various spectroscopic tecniques: electronic, Laser Magnetic Resonance (LMR),
Microwave Optical Double Resonance (MODR), millimeter-wave (mmW) and Fourier-Transform Far InfraRed (FT-FIR) spectroscopy.
Briefly, the rotational constants were first derived from the electronic spectrum \citep{Dre59}, while quartic centrifugal 
distortion and fine interaction parameters were obtained by means of LMR studies \citep{Hil79}.
Later on, \citet{Coo83} determined the hyperfine coupling constants using MODR spectroscopy and \citet{Kan91} measured the 
pure rotational spectrum in the range 265 - 531 GHz using a submillimeter-wave spectrometer.
Lastly, \citet{Mor97} observed the far-infrared spectrum in the 102 - 265 cm$^{-1}$ region with a high-resolution FT spectrometer.

In the present work, newly observed rotational transitions from 588 GHz to 1.13 THz have been measured for ND$_{2}$ with microwave accuracy.
A global analysis including also all previous field-free rotational data has been performed thus obtaining a considerably 
improved set of spectroscopic constants, from which more accurate predictions of transition frequencies of ND$_{2}$ can be 
produced in a wide range of its rotational spectrum.

\begin{figure*}[t]
	\centering
	\includegraphics{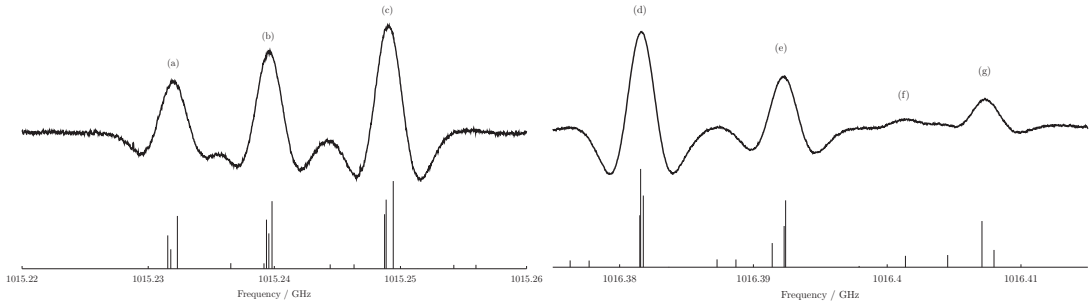}
	\caption{
		Observed and predicted frequencies for two fine components of the $N=3_{13}\gets2_{02}$ transition at 1015-1016 GHz. Both scans were recorded with a RC constant of 3 ms and a total integration time of about 200 s. \emph{Left panel}: $J=3.5\gets2.5$ (a) ($F_1=2.5\gets1.5$, $F=2.5\gets1.5$),($F_1=2.5\gets1.5$, $F=1.5\gets0.5$),($F_1=2.5\gets$1.5, $F=3.5\gets2.5$) (b) ($F_1=3.5\gets2.5$, $F=3.5\gets2.5$)($F_1=3.5\gets$2.5, $F=2.5\gets1.5$)($F_1=3.5\gets2.5$, $F=4.5\gets3.5$) (c) ($F_1=4.5\gets3.5$, $F=3.5\gets2.5$)($F_1=4.5\gets3.5$, $F=4.5\gets3.5$)($F_1=4.5\gets3.5$, $F=5.5\gets4.5$) \emph{Right panel}: $J=2.5\gets1.5$ (d) ($F_1=3.5\gets2.5$, $F=2.5\gets1.5$),($F_1=3.5\gets4.5$, $F=2.5\gets3.5$),($F_1=3.5\gets2.5$, $F=3.5\gets2.5$) (e)($F_1=2.5\gets1.5$, $F=1.5\gets0.5$),($F_1=2.5\gets1.5$, $F=2.5\gets1.5$),($F_1=2.5\gets1.5$, $F=3.5\gets2.5$) (f) ($F_1=1.5\gets0.5$, $F=1.5\gets1.5$) (g) ($F_1=1.5\gets0.5$, $F=2.5\gets1.5$),($F_1=1.5\gets0.5$, $F=1.5\gets0.5$)
	}\label{fig-1}
\end{figure*}

\section{Experiment}
The rotational spectrum of ND$_{2}$ radical has been investigated in selected frequency regions between 588 GHz and 1.13 THz using a
source-modulation millimeter/sub-millimeter wave spectrometer, as in \citet{Dor17} and \citet{Biz16}.
The main sources of radiation were several Gunn diodes (Radiometer Physics GmbH, J. E. Carlstrom Co) which emit 
in the 80-134 GHz frequency range. Higher frequencies were obtained by using passive doublers or triplers (RPG) in cascade.
The sources were phase-locked to the suitable harmonic of a frequency synthesizer (Schomandl ND 1000) referenced to an external 
rubidium frequency standard. The frequency modulation of the radiation was obtained by sine-wave modulating at 6 kHz the reference 
signal of the wide-band Gunn synchronizer.
The absorption cell (3.25 m long, 5 cm in diameter) was equipped with two cylindrical hollow electrodes 25 cm in length at either end, and was 
wound with a plastic pipe for liquid nitrogen circulation.
A liquid-helium-cooled InSb detector (QMC Instr. Ltd. type QFI/2) was used for detection. The radiation was phase-sensitively detected 
by a lock-in amplifier at twice the modulation frequency, so that the second derivative of the spectrum profile was displayed.

The ND$_{2}$ radical was produced with an electrical glow-discharge system. A mixture of ND$_{3}$ (5-7 mTorr) and Ar (20 mTorr) 
was used, as in \citet{Kan91}.
Typically, a discharge current of 70 mA and a voltage of about 1 kV were maintained between the electrodes while recording the spectrum.
The absorption cell was cooled continuously just above the freezing point of ammonia with liquid nitrogen, in order to avoid an excessive 
overheating of the two electrodes.
The estimated uncertainties of the measurements are 50 kHz in most cases, 100 kHz in a few cases. The accuracy depends on the signal 
to noise ratio of the observed spectrum and on line-blending due to unresolved hyperfine components.

\section{Results and Analysis}
The ND$_{2}$ radical in its electronic ground state $X^{2}B_{1}$ is an asymmetric rotor of $C_{2v}$ symmetry whose dipole moment (1.82 D)
lies along the \emph{b} axis \citep{Bro79}.
Because of the spin of the unpaired electron ($S=\frac{1}{2}$) and the spin of the $^{14}$N nucleus ($I_N=1$), this radical shows a complex
spectrum with fine and hyperfine structures (see Figure~\ref{fig-1}).
In fact, each rotational level with quantum number $N\neq0$ is split into two sublevels with $J=N+\frac{1}{2}$ and $J=N-\frac{1}{2}$
because of the coupling of the rotation with the electron spin.
Each of these two sublevels is then split by the interaction with the nuclear spin of $^{14}$N. A further splitting of the nitrogen hyperfine
components is due to deuterium nuclear spin. Deuterium is a boson ($I_D=1$), therefore the total wavefunction has to be symmetric upon
exchange of the two identical nuclei, that is of $A$ symmetry. Since the $X^{2}B_{1}$ electronic state
is antisymmetric and the vibrational ground state is symmetric, then  symmetric rotational levels ($K_{-1}K_{1}=(ee)$ or $(oo)$) 
combine with antisymmetric deuterium spin functions, with $I_{D}=I_{D_1}+I_{D_2} = 1$, while antisymmetric rotational levels 
($K_{-1}K_{1}=(eo)$ or $(oe)$) combine with spin functions with $I_D= 0, 2$, which are symmetric upon exchange of the D nuclei.
Thus the symmetric rotational levels are split in three further hyperfine sublevels and the antisymmetric ones in six sublevels.
Therefore the coupling scheme used in the present work is the following:
\begin{equation*}\begin{split}
   \hat{J} &= \hat{N}+\hat{S} \\
   \hat{F}_{1} &= \hat{J}+\hat{I}_{N} \\
   \hat{F} &= \hat{F}_{1}+\hat{I}_{D}.\\
\end{split}\end{equation*}

The effective Hamiltonian for the ND$_2$ radical can be expressed as~\citep{Kan91,Mor97}:
\begin{equation}
\hat{H}=\hat{H}_{rot}+\hat{H}_{fs}+\hat{H}_{hfs},
\end{equation}
where $\hat{H}_{rot}$ is the Watson $A$-reduced Hamiltonian in the $I^{r}$ representation \citep{Wat77}, 
which includes the rotational energy and the centrifugal distortion terms. 
The fine-structure Hamiltonian $\hat{H}_{fs}$ contains the electron spin-rotation terms with constants 
$\epsilon_{ii}$ and their centrifugal dependences. The $\hat{H}_{hfs}$ operator is the hyperfine-structure 
Hamiltonian and can be separated into two components:

\begin{equation}
\hat{H}_{hfs}=\hat{H}_{hfs}(N)+\hat{H}_{hfs}(D).
\smallskip
\end{equation}

In the present analysis, the effect of isotropic ($a_{F}$) and anisotropic ($T_{ii}$) electronic spin-nuclear spin coupling constants 
were considered for both nuclei, while electric quadrupole ($eQq$) and spin-rotation ($C_{ii}$) coupling constants 
could be determined only for the $^{14}$N-nitrogen nucleus.

\begin{deluxetable*}{llDDD}[t]
	\centering
	\tabletypesize{\scriptsize}
	\tablecaption{Spectroscopic parameters derived for ND$_{2}$\label{tbl-1}}
	\tablewidth{0pt}
	\tablehead{
		\multicolumn{2}{c}{Constants}  & \multicolumn{2}{c}{Present work} & \multicolumn{2}{c}{Previous MW\tablenotemark{a}} & \multicolumn{2}{c}{Previous FIR\tablenotemark{b}}
	}
	\decimals
	\startdata
	$          A   $ & /MHz & 399989.5534(87)\tablenotemark{c}    & 399985.879( 81)& 399993.92(189)  \\
	$          B   $ & /MHz & 194498.1916(150)     & 194488.65( 16) & 194498.10(102)   \\
	$          C   $ & /MHz & 128610.4447(145)     & 128613.987( 57)& 128610.00(126)   \\
	\multicolumn{8}{l}{Centrifugal distortion} \\
	$ \Delta_{N}   $ & /MHz &      7.86074(33)     &     7.323(10) &     7.8392(108)   \\
	$ \Delta_{NK}  $ & /MHz &    -33.48376(259)    &   -33.812(25) &   -33.388(42)     \\
	$ \Delta_{K}   $ & /MHz &     198.7064(57)     &   196.771(33) &   198.783(84)     \\
	$ \delta_{N}   $ & /MHz &     3.080801(201)    & 3.03\tablenotemark{d}  &     3.0858(36)  \\
	$ \delta_{K}   $ & /MHz &       8.5567(82)     &     6.300(39) &     8.2788(294)    \\
	$    \Phi_{N}  $ & /kHz &       1.4776(80)     &       .       &     1.376(35)    \\
	$   \Phi_{NNK}  $ & /kHz &       -6.277(306)    &       .       &    -5.606(35)    \\
	$  \Phi_{KKN}   $ & /kHz &       -43.70(98)     &       .       &    -43.45(96)    \\
	$  \Phi_{K}    $ & /kHz &       320.34(90)     &       .       &     321.0(23)    \\
	$   \phi_{N}   $ & /kHz &       0.7224(33)     &       .       &     0.7402(140)  \\
	$   \phi_{NK}  $ & /kHz &       -0.834(99)     &       .       &    -1.01(21)     \\
	$   \phi_{K}   $ & /kHz &       65.90(133)     &       .       &     33.37(65)    \\
	$    L_{KKKN}   $ & /kHz &      0.08336(141)    &       .       &     0.0707(38)   \\
	$     L_{K}    $ & /kHz &      -0.7022(85)     &       .       &    -0.694(26)    \\
	$      l_{K}   $ & /kHz &      -0.1230(33)     &       .       &    -0.0311(16)   \\
	$      M_{K}   $ & /Hz  &        1.069(38)     &       .       &     1.08(10)     \\
	\multicolumn{8}{l}{Fine interaction} \\
	$\epsilon_{aa} $ & /MHz &     -5128.1262(215)  & -5127.81(11)    & -5127.71(41)     \\
	$\epsilon_{bb} $ & /MHz &      -668.6973(180)  &  -668.507(78)   &  -668.759(160)   \\
	$\epsilon_{cc} $ & /MHz &         3.4616(166)  &     3.413(63)   &     3.23(12)     \\
	$\Delta^{S}_{N} $ & /MHz &       0.07613(57)   &     0.0469(66)  &     0.0779(29)   \\
	$\Delta^{S}_{KN+NK}$ & /MHz &    -0.9498( 57)  &    -0.967(81)   &    -0.875(29)    \\
	$\Delta^{S}_{NK} $ & /MHz &        1.045(201)  &      .          &            .     \\
	$\Delta^{S}_{K} $ & /MHz &        9.8548(76)   &     9.587(66)   &     9.476(99)    \\
	$\delta^{S}_{N} $ & /MHz &       0.03843(36)   &      .          &     0.0353(16)   \\
	$\delta^{S}_{K} $ & /MHz &        0.1005(59)   &     0.0356(45)  &     0.150(36)    \\
	$\Phi^{S}_{NKK} $ & /kHz &          3.76(51)   &      .          &           .      \\
	$\Phi^{S}_{K}   $ & /kHz &        -23.10(81)   &      .          &     -15.36(72)   \\
	\multicolumn{8}{l}{Hyperfine interaction} \\
	$   a_{F}(N)    $ & /MHz &     28.0577(122)    &  28.055(33)     &     .            \\
	$     T_{aa}(N) $ & /MHz &    -43.1451(183)    & -43.136(48)     &     .            \\
	$     T_{bb}(N) $ & /MHz &    -44.2715(213)    & -44.277(63)     &     .            \\
	$     C_{aa}(N) $ & /MHz &      0.2660(75)     &   0.269(27)     &     .            \\
	$     C_{bb}(N) $ & /MHz &      0.0458(55)     &   0.045(21)     &     .            \\
	$     C_{cc}(N) $ & /MHz &      0.0145(64)     &   0.019(22)     &     .            \\
	$     a_{F}(D)  $ & /MHz &    -10.2446(103)    & -10.241(28)     &     .            \\
	$     T_{aa}(D) $ & /MHz &      2.8536(168)    &   2.874(45)     &     .            \\
	$     T_{bb}(D) $ & /MHz &     -2.0916(247)    &  -2.108(69)     &     .            \\
	$     X_{aa}(N) $ & /MHz &      0.2191(297)    &  0.213(84)      &     .            \\
	$     X_{bb}(N) $ & /MHz &      -3.744(34)     &  -3.75(11)      &     .            \\
	\hline
	\multicolumn{6}{l}{Number of FIR lines} & \multicolumn{2}{r}{181} \\
	\multicolumn{6}{l}{Standard deviation of the FIR data /cm\textsuperscript{-1}} & \multicolumn{2}{r}{$6.9\times 10^{-4}$} \\
	\multicolumn{6}{l}{Number of MODR lines} & \multicolumn{2}{r}{198} \\
	\multicolumn{6}{l}{Standard deviation of the MODR data /MHz} & \multicolumn{2}{r}{0.376} \\
	\multicolumn{6}{l}{Number of MW lines} & \multicolumn{2}{r}{182} \\
	\multicolumn{6}{l}{Standard deviation of the MW data /kHz} & \multicolumn{2}{r}{51.0} \\
	\multicolumn{6}{l}{Fit standard deviation} & \multicolumn{2}{r}{0.87} \\
	\multicolumn{6}{l}{$N'_{max}$, ${K'_a}_{ max}$} & \multicolumn{2}{r}{13, 10} \\
	\enddata
	\centering
	\tablenotetext{a}{\citet{Kan91}} 
	\tablenotetext{b}{\citet{Mor97}. The values, originally reported in cm\textsuperscript{-1}, are converted to MHz.}
	\tablenotetext{c}{Values in parenthesis denote one standard deviation and apply to the last digits.}
	\tablenotetext{d}{Fixed}
\end{deluxetable*}

The initial predictions were made using the Pickett's program SPCAT \citep{Pic91} with the constants derived by \citet{Kan91}.
The production of ND$_{2}$ was optimized by observing the $J=0\gets0$ component of the $N=1_{10}\gets1_{01}$ transition.
Then, the fine components of new eight rotational transitions were recorded and their hyperfine structure could be easily assigned
thanks to the already known hyperfine-interaction parameters.
Sixty-four distinct frequencies were measured and analyzed in a weighted least-squares procedure, in which previous data 
(millimeter/sub-millimeter, MODR and FIR studies) were included.
In this procedure, performed with the Pickett's program SPFIT \citep{Pic91}, the mmW and sub-mmW lines previously measured 
by \citet{Kan91} were given uncertainties of 50 kHz, whereas the FIR data from \citet{Mor97} were given uncertainties of 
0.0007 cm\textsuperscript{-1}, as in the analysis of NH$_{2}$ \citep{Mul99}.
For the MODR data \citep{Coo83} the same uncertainties given in the original work were assumed (typically 0.5 MHz).
Surprisingly, the older sets of frequency data were never analyzed simultaneously, so that different sets of spectroscopic constants, 
not fully consistent, have been reported in the literature for ND$_{2}$. The present global analysis allows to obtain spectroscopy 
constants which are compatible with the entire body of rotational data available for this radical. Moreover, we were able to 
complete and/or correct previous assignments. As far as MODR data are concerned, about 20\% of the observed transitions were not considered
in the fit of \citet{Coo83} because  of the failure to unambiguously assign their hyperfine components. Our more extensive 
data analysis allowed us to overcome this difficulty, so that a large number of previously unused MODR frequencies could be 
included in our fit. Furthermore, during the merging process some FIR transitions appeared to be misassigned, so they have been 
reassigned correctly in this study.
In particular, a few transitions with $N \ge 10$ had been wrongly assigned as $\Delta K_{a}=\pm3$ instead of $\Delta K_{a}=\pm1$.
In the final fit, unresolved lines were treated using the intensity-weighted average of the individual components involved.
A total number of 41 spectroscopic constants were included in the least squares analysis, giving a satisfactory dimensionless standard
deviation of 0.87.
Table~\ref{tbl-1} shows the list of spectroscopic parameters determined in the present work, where they are compared to those 
obtained previously. In general, all derived parameters are determined with higher precision than in the previous 
works: for instance, the standard errors of the rotational constants are one order of magnitude lower and the hyperfine constants 
are three times more precise. This is mainly due to the simultaneous analysis of MW spectroscopy measurements
(high frequency precision and resolving power) with FT-FIR data (large range of $N$ and $K_a$ quantum numbers).
The newly determined values of the spectroscopic constants agree well with those previously reported by \citet{Kan91} and \citet{Mor97},
the only exceptions being $\phi_K$ and $l_K$. This discrepancy should be ascribed to the reassignment process of some FIR transitions (see above).

It is well known that the lightness and non-rigidity of triatomic hydrides of $C_{2v}$ symmetry (e.g. CH$_2$, H$_2$O, NH$_2$) cause abnormal
centrifugal distortion effects \citep{Bru05,Yu12,Mar14}, so that describing the rotational energy levels using a Watson-type Hamiltonian presents
some difficulties due to the slow convergence of the standard power series expansion of the rotational angular momentum operators.
As a matter of fact, a huge number of centrifugal distortion terms (mainly $K_a$ dependent) has to be included in the Hamiltonian model in order to
reproduce highly accurate rotational data within their experimental uncertainty.
This difficulties are less pronounced for the heavier doubly deuterated isotopologues, for which the centrifugal distortion effects are smaller.
As far as the Amidogen radical is concerned, on the average the quartic centrifugal distortion constants determined for ND$_2$ are ca. three times
smaller than those of NH$_2$, and the sextic ones ca. seven times smaller~\citep{Mar14}.
No problem of convergence has been experienced in the present analysis using the standard semi-rigid Hamiltonian, so that the experimental transition
frequencies could be satisfactorily reproduced ($\sigma=0.87$) using, as fit parameters, 16 centrifugal distortion constants less than in the case
of NH$_2$ \citep{Mar14}.
For this reason no other form of the rotational Hamiltonian was considered for the analysis.

Table~\ref{tbl-2} provides an example of the machine-readable table included as 
Electronic Supplementary Information. Files used with the SPFIT/SPCAT program suite are also included as supplementary materials.
These files can be used to predict the rotational spectrum at a given temperature and in selected frequency regions.
It should be noted that the prediction of transition frequencies corresponding to quantum numbers higher than those involved in the present analysis
(i.e. $N>$ 13 and $K_{a}>$ 10) could be affected by systematic errors due to the truncation of the slowly convergent power series expansion of the
standard semi-rigid Hamiltonian.
It is not easy to make an {\it a priori} evaluation of the magnitude of these model-dependent errors, however an estimate of the convergence radius
of a power series can provide useful information \citep{Bru05}.
For instance, considering the series of the purely $K_{a}$ dependent centrifugal distortion terms, $\sum_n c_n \left(K_{a}^2\right)^n$,
its convergence radius is the inverse square-root of the absolute value of $\lim_{n\to\infty}{c_{n}/c_{n-1}}$, which can be estimated from the
intercept of the plot of $c_{n}/c_{n-1}$ versus $1/n$. For the present centrifugal analysis, a convergence radius of $K_a = 17$ results:
at this value of $K_a$, $|c_{n-1}|\left(K_{a}^2\right)^{n-1}\sim |c_{n}|\left(K_{a}^2\right)^n$ holds,
therefore the missing higher order term gives a substantial contribution to the energy of the level.
In conclusion, the semi-rigid Hamiltonian with the present set of constants if fully accurate to predict the ND$_2$ spectrum for $K_{a}\leqslant 10$,
it is still appropriate, with less accuracy, for $10 < K_{a}\leqslant 17$, but for $K_a > 17$ no reliable predictions of the spectrum can be obtained.

\section{Conclusion}
The rotational spectrum of the doubly deuterated variant of Amidogen radical ND$_{2}$ in its ground electronic state $X^{2}B_{1}$ 
has been investigated in the frequency region 588 - 1131 GHz. New measurements have been analyzed in a global fit including 
previous mmW, sub-mmW, MODR and FIR data. Some misassignments or incomplete assignments of previous data sets have been corrected, 
so that it has been possible to obtain a set of very accurate spectroscopy constants which are compatible with the entire body of 
rotational data presently available for this radical. Accurate predictions are now available up to 8 THz for those transitions 
with $N\leqslant13$ and $K_{a}\leqslant10$. This work provides a facility set of data for the radio-observation of fully deuterated Amidogen radical.

\begin{deluxetable*}{ccccccccccccccDDDDc}[t]
	\tabletypesize{\scriptsize}
	\tablecaption{Observed frequencies and residual from the final fit for ND$_{2}$ Amidogen radical\label{tbl-2}}
	\tablewidth{0pt}
	\tablehead{
		\colhead{$N'$} & \colhead{$K'_a$} & \colhead{$K'_c$} & \colhead{$J'$} & \colhead{$F_{1}'$} & \colhead{$I_{D}'$} & \colhead{$F'$} &
		\colhead{$N$} & \colhead{$K_a$} & \colhead{$K_c$} & \colhead{$J$} & \colhead{$F_{1}$} & \colhead{$I_{D}$} & \colhead{$F$} & 
		\multicolumn{2}{c}{Obs. Frequency} & \multicolumn{2}{c}{Uncert.} & \multicolumn{2}{c}{Obs.-Calc.} & 
		\multicolumn{2}{c}{Relative weight\tablenotemark{a}} & \colhead{Reference} \\
		\colhead{} &\colhead{} &\colhead{} & \colhead{} & \colhead{} & \colhead{} &\colhead{} & 
		\colhead{} &\colhead{} & \colhead{} & \colhead{}  &\colhead{} &\colhead{} & \colhead{} &
		\multicolumn{2}{c}{(MHz)} & \multicolumn{2}{c}{(MHz)} & \multicolumn{2}{c}{(MHz)} & \multicolumn{2}{c}{} & \colhead{} \\
	}
	\startdata
	3 & 1 & 3 & 4 & 3 & 1 & 3 & 2 & 0 & 2 & 3 & 2 & 1 & 2 &  \multicolumn{2}{c}{1015231.991} & \multicolumn{2}{c}{0.050} & \multicolumn{2}{c}{ 0.016} & \multicolumn{2}{c}{0.3157} & This work \\
	3 & 1 & 3 & 4 & 3 & 1 & 2 & 2 & 0 & 2 & 3 & 2 & 1 & 1 &  \multicolumn{2}{c}{1015231.991} & \multicolumn{2}{c}{0.050} & \multicolumn{2}{c}{ 0.016} & \multicolumn{2}{c}{0.1845} & This work \\
	3 & 1 & 3 & 4 & 3 & 1 & 4 & 2 & 0 & 2 & 3 & 2 & 1 & 3 &  \multicolumn{2}{c}{1015231.991} & \multicolumn{2}{c}{0.050} & \multicolumn{2}{c}{ 0.016} & \multicolumn{2}{c}{0.4998} & This work \\
	3 & 1 & 3 & 4 & 4 & 1 & 4 & 2 & 0 & 2 & 3 & 3 & 1 & 3 &  \multicolumn{2}{c}{1015239.596} & \multicolumn{2}{c}{0.050} & \multicolumn{2}{c}{-0.023} & \multicolumn{2}{c}{0.3230} & This work \\
	3 & 1 & 3 & 4 & 4 & 1 & 3 & 2 & 0 & 2 & 3 & 3 & 1 & 2 &  \multicolumn{2}{c}{1015239.596} & \multicolumn{2}{c}{0.050} & \multicolumn{2}{c}{-0.023} & \multicolumn{2}{c}{0.2326} & This work \\
	3 & 1 & 3 & 4 & 4 & 1 & 5 & 2 & 0 & 2 & 3 & 3 & 1 & 4 &  \multicolumn{2}{c}{1015239.596} & \multicolumn{2}{c}{0.050} & \multicolumn{2}{c}{-0.023} & \multicolumn{2}{c}{0.4445} & This work \\
	3 & 1 & 3 & 4 & 5 & 1 & 4 & 2 & 0 & 2 & 3 & 4 & 1 & 3 &  \multicolumn{2}{c}{1015249.043} & \multicolumn{2}{c}{0.050} & \multicolumn{2}{c}{-0.023} & \multicolumn{2}{c}{0.2580} & This work \\
	3 & 1 & 3 & 4 & 5 & 1 & 5 & 2 & 0 & 2 & 3 & 4 & 1 & 4 &  \multicolumn{2}{c}{1015249.043} & \multicolumn{2}{c}{0.050} & \multicolumn{2}{c}{-0.023} & \multicolumn{2}{c}{0.3270} & This work \\
	3 & 1 & 3 & 4 & 5 & 1 & 6 & 2 & 0 & 2 & 3 & 4 & 1 & 5 &  \multicolumn{2}{c}{1015249.043} & \multicolumn{2}{c}{0.050} & \multicolumn{2}{c}{-0.023} & \multicolumn{2}{c}{0.4150} & This work \\
	\enddata
	\tablecomments{Table 2 is published in its entire form in the electronic edition of the ApJS.
		A portion is shown here for guidance regarding its form and content.}
	\tablenotetext{a}{For blended transitions}
\end{deluxetable*}

\section{Acknowledgments}
This study was supported by Bologna University (RFO funds) and by MIUR (Project PRIN 2015: STARS in the CAOS, Grant Number 2015F59J3R).

\end{document}